\documentclass[reprint,superscriptaddress,nofootinbib,aps,twocolumn]{revtex4}
\usepackage{soul}
\usepackage{bm}
\usepackage[T1]{fontenc}
\usepackage{amsmath,amsfonts,bbm, graphicx, color}
\usepackage{amssymb}
\usepackage{appendix}
\usepackage{mathtools}
\usepackage{dsfont}
\usepackage{float}
\usepackage{relsize}
\usepackage{xcolor}
\usepackage{tikz}
\usetikzlibrary{svg.path}
\usepackage{soul}

\definecolor{orcidlogocol}{HTML}{A6CE39}
\tikzset{
  orcidlogo/.pic={
    \fill[orcidlogocol] svg{M256,128c0,70.7-57.3,128-128,128C57.3,256,0,198.7,0,128C0,57.3,57.3,0,128,0C198.7,0,256,57.3,256,128z};
    \fill[white] svg{M86.3,186.2H70.9V79.1h15.4v48.4V186.2z}
                 svg{M108.9,79.1h41.6c39.6,0,57,28.3,57,53.6c0,27.5-21.5,53.6-56.8,53.6h-41.8V79.1z M124.3,172.4h24.5c34.9,0,42.9-26.5,42.9-39.7c0-21.5-13.7-39.7-43.7-39.7h-23.7V172.4z}
                 svg{M88.7,56.8c0,5.5-4.5,10.1-10.1,10.1c-5.6,0-10.1-4.6-10.1-10.1c0-5.6,4.5-10.1,10.1-10.1C84.2,46.7,88.7,51.3,88.7,56.8z};
  }
}

\newcommand\orcid[1]{\href{https://orcid.org/#1}{\mbox{\scalerel*{
\begin{tikzpicture}[yscale=-1,transform shape]
\pic{orcidlogo};
\end{tikzpicture}
}{|}}}}
\usepackage[colorlinks]{hyperref}
\usepackage[hyphenbreaks]{breakurl}

\hypersetup{
	bookmarksnumbered,
	pdfstartview={FitH},
	citecolor={blue},
	linkcolor={darkblue},
	urlcolor={blue},
	pdfpagemode={UseOutlines}}
\definecolor{darkgreen}{RGB}{20,100,20}
\definecolor{darkblue}{RGB}{0,0,130}
\definecolor{darkred}{rgb}{.8,0,0}

\usepackage{soul}
		
\usepackage{colonequals}

\begin{document}

\title{Relativity of superluminal observers in $1+3$ spacetime}

\author{Andrzej Dragan}
\email{dragan@fuw.edu.pl}
\affiliation{Institute of Theoretical Physics, University of Warsaw, Pasteura 5, 02-093 Warsaw, Poland}
\affiliation{Centre for Quantum Technologies, National University of Singapore, 3 Science Drive 2, 117543 Singapore, Singapore}

\author{Kacper Dębski}
\affiliation{Institute of Theoretical Physics, University of Warsaw, Pasteura 5, 02-093 Warsaw, Poland}

\author{Szymon Charzyński}
\affiliation{Department of Mathematical Methods in Physics, University of Warsaw, Pasteura 5, 02-093 Warsaw, Poland}

\author{Krzysztof Turzyński}
\affiliation{Institute of Theoretical Physics, University of Warsaw, Pasteura 5, 02-093 Warsaw, Poland}

\author{Artur Ekert}
\affiliation{Centre for Quantum Technologies, National University of Singapore, 3 Science Drive 2, 117543 Singapore, Singapore}
\affiliation{Mathematical Institute, University of Oxford, Woodstock Road, Oxford OX2 6GG, United Kingdom}
\affiliation{Okinawa Institute of Science and Technology, Onna, Okinawa 904-0495, Japan}

\begin{abstract}We develop an extension of special relativity in $1+3$ dimensional spacetime to account for superluminal inertial observers and show that such an extension rules out the conventional dynamics of mechanical point-like particles and forces one to use a field-theoretic framework. Therefore we show that field theory can be viewed as a direct consequence of extended special relativity.
\end{abstract}

\maketitle

\section{Introduction}

In a recent paper Dragan and Ekert showed that extending special relativity to account for superluminal particles and observers does not result in causal paradoxes \cite{Dragan2020}, as was commonly believed \cite{Tolman1917}. Instead, such an extension modifies the notion of causality in the same way quantum theory does. In particular, it was shown that when Galilean relativity (involving {\em all} inertial observers) is assumed in $1+1$ dimensional spacetime, indeterministic behavior and motion along multiple paths analogous to quantum mechanical superpositions becomes inevitable. This result concluded the debate, initiated by Tolman \cite{Tolman1917}, about alleged problems with causality triggered by the presence of superluminal particles in relativity \cite{Pirani1970,Parmentola1971,Landsberg1972,Goldhaber1975,Barrowes1977,Basano1977,Maund1979,Basano1980,Antippa1972}.

However, generalizing this scheme to a relativistic framework in a $1+3$ dimensional spacetime poses some serious challenges, both mathematical and interpretational \cite{Parker1969}. The main obstacle, pointed out by Marchildon, Antippa and Everett \cite{Machildon1983}, results from the fact that the smallest group containing superluminal transformations is $SL(4,\mathbb{R})$. This group, however, cannot be a symmetry group, because it contains elements such as direction-dependent dilations, which are not observed. In this work, we propose a way to overcome these difficulties and consistently extend special relativity in $1+3$ dimensional spacetime to superluminal frames of reference. We argue that within such an extension the standard classical dynamics of a point-like particle cannot be supported and the only relativistically invariant dynamics of any physical systems involving both subluminal and superluminal observers requires a field-theoretic framework.

\section{Kinematics}

An orthodox $1+3$ dimensional subluminal Lorentz boost between two mutually unrotated frames $(ct,\boldsymbol{r})$ and $(ct',\boldsymbol{r'})$ moving with a relative velocity $\boldsymbol{V}$ leaves the transversal spacial components of four-position unaffected: $\boldsymbol{r'} -\frac{\boldsymbol{r'}\cdot\boldsymbol{V}}{V^2}\boldsymbol{V} = \boldsymbol{r}-\frac{\boldsymbol{r}\cdot\boldsymbol{V}}{V^2}\boldsymbol{V}$. The longitudinal component undergoes the standard Lorentz transformation: $\frac{\boldsymbol{r'}\cdot\boldsymbol{V}}{V^2}\boldsymbol{V} = \frac{\frac{\boldsymbol{r}\cdot\boldsymbol{V}}{V^2}-t}{\sqrt{1-\frac{V^2}{c^2}}}\boldsymbol{V}$. Adding these two equations by sides and supplementing them with the Lorentz transformation for the temporal coordinate $t'$ leads to the $1+3$ dimensional Lorentz boost between a pair of mutually unrotated observers \cite{Dragan2021}:
\begin{eqnarray}
\label{lorentz}
\boldsymbol{r'} &=& \boldsymbol{r}-\frac{\boldsymbol{r}\cdot\boldsymbol{V}}{V^2}\boldsymbol{V} + \frac{\frac{\boldsymbol{r}\cdot\boldsymbol{V}}{V^2}-t}{\sqrt{1-\frac{V^2}{c^2}}}\boldsymbol{V},\nonumber \\
ct'  &=& \frac{ct-\frac{\boldsymbol{r}\cdot\boldsymbol{V}}{c}}{\sqrt{1-\frac{V^2}{c^2}}}.
\end{eqnarray}
Such a boost preserves the metric, so that the form of the spacetime interval remains the same in both frames: $c^2\text{d}t^2 - \text{d}\boldsymbol{r}\cdot\text{d}\boldsymbol{r} = c^2\text{d}t'^2 - \text{d}\boldsymbol{r'}\cdot\text{d}\boldsymbol{r'}$.
The inverse transformation can be obtained by substituting $\boldsymbol{V}\to-\boldsymbol{V}$, as well as $\boldsymbol{r}\leftrightarrow\boldsymbol{r'}$ and $t\leftrightarrow t'$, instead of algebraically reversing the equations. A velocity transformation formula is obtained by evaluating the derivative $\boldsymbol{v'}\equiv \frac{\text{d}\boldsymbol{r'}}{\text{d}t'}$:
\begin{align}
\label{veltrans}
\boldsymbol{v'} = \frac{\sqrt{1-\frac{V^2}{c^2}}\left(\boldsymbol{v}-\frac{\boldsymbol{v}\cdot\boldsymbol{V}}{V^2}\boldsymbol{V}\right) 
- \left(\boldsymbol{V}-\frac{\boldsymbol{v}\cdot\boldsymbol{V}}{V^2} \boldsymbol{V} \right)}{1-\frac{\boldsymbol{v}\cdot\boldsymbol{V}}{c^2}}.
\end{align}

Let us now carry out a similar procedure for the superluminal transformations with $V>c$ discussed in \cite{Dragan2020}. Superluminal observers have been discussed since the 1960s \cite{Sudarshan1962}, but even the simplest $1+1$ dimensional transformations considered in literature were missing the anti-symmetric factor \cite{Recami1974,Recami1986} rendering the whole theory non-covariant. The correct expressions containing that anti-symmetric term appeared for the first time in a paper by Parker \cite{Parker1969}:
\begin{align}
\label{super1d}
x' &= \pm\frac{V}{|V|}\frac{x-Vt}{\sqrt{\frac{V^2}{c^2}-1}}, \nonumber \\
t' &= \pm\frac{V}{|V|}\frac{t-\frac{V}{c^2}x}{\sqrt{\frac{V^2}{c^2}-1}}.
\end{align}
This transformation changes the sign of the metric: $c^2\text{d}t^2 - \text{d}x^2 = - c^2\text{d}t'^2 + \text{d}x'^2$, but preserves null intervals, so the speed of light is the same in both subluminal and superluminal frames of reference, as required. The unspecified sign $\pm$ cannot be uniquely determined, as no $V\to 0$ exists, but, as discussed in \cite{Dragan2020}, we will pick the negative sign convention so that the transformation \eqref{super1d} remains a hyperbolic rotation.

In order to generalize \eqref{super1d} to the $1+3$ dimensional spacetime we introduce an additional pair of spatial dimensions $y$ and $z$ for the "resting" observer, as well as an extra pair $\xi'$ and $\chi'$ for the superluminal one. We will also assume that the "transversal" components  of the four-position are not affected by the superluminal transformation. It follows that for the motion along the $x$ axis with a superluminal velocity the spacetime interval undergoes the following transformation: $c^2\text{d}t^2 - \text{d}\boldsymbol{r}\cdot\text{d}\boldsymbol{r} = -c^2\text{d}t'^2 + \text{d}x'^2 - \text{d}\xi'^2 -\text{d}\chi'^2$, so the spacetime metric is affected, but the null intervals are still preserved. This result indicates that the laws of physics in the inertial superluminal frame of reference are different from those within the orthodox family of subluminal frames. This agrees with the argument put forth by Machildon {\em et al.} \cite{Machildon1983}. Furthermore, Sutherland and Shepanski argue \cite{Sutherland1986} that this is due to the fact the the "spatial" component of the four-position in the superluminal frame, defined as $(x',\xi',\chi')$, characterizes a non-Euclidean space. In this work we follow a different interpretation first mentioned by Demers \cite{Demers1975} and developed in \cite{Dragan2020}: the signs of the metric components indicate that the extra pair of dimensions $\xi'$ and $\chi'$ are in fact temporal, because they enter the metric with the same sign as the temporal coordinate $t'$. According to this interpretation a superluminal frame of reference is equipped with a single spatial dimension that we will denote with $r'$, as well as three temporal dimensions $\boldsymbol{t'}$. In this paper, we will investigate physical consequences of such a hypothesis.   

Let us consider a subluminal frame $(ct,\boldsymbol{r})$ and an unrotated, superluminal frame $(c\boldsymbol{t'},r')$ moving with a superluminal velocity $\boldsymbol{V}$. Here, by "unrotated" we mean that the resting frame should move relative to the superluminal frame with velocity $-\boldsymbol{V}$, so that the inverse transformation leads to a sign flip in relative velocity. First off, we note that the superluminal transformation apparently leaves the transversal spatial components of four-position unaffected: $c\boldsymbol{t'} -c\frac{\boldsymbol{t'}\cdot\boldsymbol{V}}{V^2}\boldsymbol{V} = \boldsymbol{r}-\frac{\boldsymbol{r}\cdot\boldsymbol{V}}{V^2}\boldsymbol{V}$. The longitudinal component along the direction $\boldsymbol{V}$ undergoes the transformation given by \eqref{super1d}: $\frac{\boldsymbol{t'}\cdot\boldsymbol{V}}{V^2}\boldsymbol{V} = -\frac{\boldsymbol{V}}{V}\frac{t-\frac{\boldsymbol{r}\cdot\boldsymbol{V}}{c^2}}{\sqrt{\frac{V^2}{c^2}-1}}$ and the spatial coordinate transforms according to $r'=-\frac{\frac{\boldsymbol{r}\cdot\boldsymbol{V}}{V}-Vt}{\sqrt{\frac{V^2}{c^2}-1}}$. Adding up the first two equations and supplementing the result with the third one yields a result first introduced by Dragan and Ekert \cite{Dragan2020,Dragan2021}:
\begin{eqnarray}
\label{super3d}
r' &=& \frac{Vt-\frac{\boldsymbol{r}\cdot\boldsymbol{V}}{V}}{\sqrt{\frac{V^2}{c^2}-1}},\nonumber \\
c\boldsymbol{t}' &=& \boldsymbol{r}-\frac{\boldsymbol{r}\cdot\boldsymbol{V}}{V^2}\boldsymbol{V} + \frac{\frac{\boldsymbol{r}\cdot\boldsymbol{V}}{Vc}-\frac{ct}{V}}{\sqrt{\frac{V^2}{c^2}-1}}\boldsymbol{V},
\end{eqnarray}
which is the coordinate transformation between a subluminal and a superluminal inertial observer in $1+3$ spacetime, we will call {\em a superboost}. The purpose of this paper is to investigate physical consequences of \eqref{super3d}.

As mentioned before, the $V\to 0$ limit of this transformation does not exist, but the infinite velocity limit does. For $V\to\infty$, the formulae \eqref{super3d} reduce to the following {\em superflip}:
\begin{eqnarray}
\label{inf3d}
r' &=&ct,\nonumber \\
c\boldsymbol{t}' &=& \boldsymbol{r},
\end{eqnarray}
regardless of the direction of velocity $\boldsymbol{V}$. The spacetime interval undergoes the following transformation under \eqref{super3d}:
\begin{align}
c^2\text{d}t^2 - \text{d}\boldsymbol{r}\cdot \text{d}\boldsymbol{r}= - c^2\text{d}\boldsymbol{t'}\cdot \text{d}\boldsymbol{t'} + \text{d}r'^2, 	
\end{align}
which does not depend on $\boldsymbol{V}$. This signifies the fact that laws of physics are the same across the whole family of superluminal inertial observers, although these laws differ from the ones common among subluminal inertial observers due to a different metric. In the following we will adopt a convention, in which the spacetime metric in subluminal frames is $\eta_{\mu \nu} \equiv \text{diag}(1,-1,-1,-1)$, and in superluminal frames it is defined as $\eta'_{\mu \nu} \equiv \text{diag}(1,1,1,-1)$, so that the four-positions $x^\mu \equiv (ct, \boldsymbol{r})$ and $x'^\mu \equiv (c\boldsymbol{t'},r')$ and all other four-vectors are related by:
\begin{align}
\label{intervaltrans}
\eta_{\mu\nu} x^\mu x^\nu = -\eta'_{\mu\nu} x'^\mu x'^\nu. 
\end{align}

The superboost \eqref{super3d} can be also rewritten in the following form:
\begin{align}
\label{super3d2}
r'  &= \frac{ct-\frac{c\boldsymbol{r}\cdot\boldsymbol{V}}{V^2}}{\sqrt{1-\frac{c^2}{V^2}}},\nonumber \\
c\boldsymbol{t}' &= \boldsymbol{r}-\frac{\boldsymbol{r}\cdot\boldsymbol{V}}{V^2}\boldsymbol{V} + \frac{\frac{\boldsymbol{r}\cdot\boldsymbol{V}}{V^2}-\frac{c^2t}{V^2}}{\sqrt{1-\frac{c^2}{V^2}}}\boldsymbol{V},
\end{align}
in which it becomes clear that it can be decomposed into the subluminal Lorentz boost \eqref{lorentz} for velocity $\boldsymbol{V}\to \frac{c^2}{V}\frac{\boldsymbol{V}}{V}$ followed by the superflip \eqref{inf3d}. The first operation takes an object moving with a superluminal speed $\boldsymbol{V}$ to a frame, in which the object moves infinitely fast, the latter goes to the rest frame of the object, in which $r'$ is constant. 
Any superboost \eqref{super3d2} can be decomposed into a pair of such operations.

In subluminal relativity the worldline of a moving observer coincides with his temporal axis $t'$ and it uniquely defines the $1+3$ decomposition of spacetime into time and space, as the three-dimensional space $\boldsymbol{r'}$ is orthogonal to the time axis $t'$. In order to uniquely define the orientation of the spacial axes $\boldsymbol{r'}$ additional information has to be provided. To supplement the missing information one usually assumes that the two moving frames are mutually unrotated, which is equivalent of saying that the inverse transformation involves a simple sign flip in relative velocity.

In the case of superluminal observers the situation is different, but some analogies remain. A trajectory of the superluminal observer does not necessarily coincide with any of the three temporal axes $\boldsymbol{t'}$. Also the one-dimensional space $r'$ of the superluminal observer is not uniquely defined by his worldline and additional information has to be provided to define it. Our derivation may be interpreted as a way to choose the $r'$ axis. The missing information is again provided by assuming that the two considered reference frames are mutually unrotated. These observations are a part of a bigger picture that will be better understood once we deal with dynamics. In particular we will show that point-like trajectories need to be rejected anyway, as incompatible with relativistic requirements involving superluminal observers.

Our interpretation that the superluminal observer characterizes spacetime using three temporal dimensions $\boldsymbol{t'}$ poses several interpretational challenges. The first question: how to even define a velocity in a frame that has more than one temporal dimension? In order to answer this question, let us determine the inverse superboost to \eqref{super3d}. It can be obtained by an algebraic reversal, which turns out to be equivalent to substituting $\boldsymbol{V}\to-\boldsymbol{V}$ and interchanging primes according to substitutions $r' \leftrightarrow ct$ and $c\boldsymbol{t'} \leftrightarrow \boldsymbol{r}$: 
\begin{eqnarray}
\label{superinverse3d}
ct  &=& \frac{\frac{Vr'}{c}+\frac{\boldsymbol{V}\cdot c\boldsymbol{t'}}{V}}{\sqrt{\frac{V^2}{c^2}-1}},\nonumber \\
\boldsymbol{r} &=& c\boldsymbol{t'}-\frac{\boldsymbol{V}\cdot c\boldsymbol{t'}}{V^2}\boldsymbol{V} + \frac{\frac{\boldsymbol{V}\cdot \boldsymbol{t'}}{V}+\frac{r'}{V}}{\sqrt{\frac{V^2}{c^2}-1}}\boldsymbol{V}.
\end{eqnarray}
The question that remains is whether $-\boldsymbol{V}$ is an actual velocity of the frame $(ct, \boldsymbol{r})$ observed in the frame $(c\boldsymbol{t'}, r')$, as required by the fact that the two frames are mutually unrotated, and how such a velocity is even defined? Let us take the point of view of the superluminal primed observer and consider the origin of the "resting", unprimed frame $\boldsymbol{r}=0$. We should expect that according to the right definition of velocity, that origin should move with velocity $-\boldsymbol{V}$. By substituting $\boldsymbol{r}=0$ into \eqref{super3d} we find:
\begin{eqnarray}
r'  &=& \frac{Vt}{\sqrt{\frac{V^2}{c^2}-1}},\nonumber \\
c\boldsymbol{t}' &=& - \frac{\frac{ct}{V} \boldsymbol{V}}{\sqrt{\frac{V^2}{c^2}-1}}.
\end{eqnarray}
Elimination of $t$ from the equations gives $r'\frac{\boldsymbol{V}}{V^2} = -\boldsymbol{t'}$. This result inspires the definition of a velocity $\boldsymbol{v'}$ of an object moving in a superluminal frame such that the following equation is satisfied:
\begin{align}
\label{supervelocityprev}
r'\frac{\boldsymbol{v'}}{v'^2} = \boldsymbol{t'}
\end{align}
and such a velocity also has the property that:
\begin{align}
\label{velproduct}
r' = \boldsymbol{v'}\cdot\boldsymbol{t'}.
\end{align}
We can now determine $v'^2$ by taking a scalar product of \eqref{supervelocityprev} with itself and plug it back into \eqref{supervelocityprev}. A rearranged differential form of \eqref{supervelocityprev} is then:
\begin{align}
\label{supervelocity}
\boldsymbol{v'} = \frac{\text{d}r'}{\text{d}t'} \frac{\text{d}\boldsymbol{t'}}{\text{d}t'},
\end{align}
where $\text{d}t' \equiv \sqrt{\text{d}\boldsymbol{t'}\cdot \text{d}\boldsymbol{t'}}$. Equation \eqref{supervelocity} will provide our definition of an instantaneous velocity in a superluminal frame.

Let us now use the definition \eqref{supervelocity} to determine the superluminal velocity composition formula. Consider an object moving in the subluminal frame with any velocity $\boldsymbol{v}$. Let us transform this velocity to the superluminal frame. By substituting $\boldsymbol{r} = \boldsymbol{v}t$ into \eqref{super3d} we obtain:
\begin{eqnarray}
r'  &=& \frac{Vt-\frac{\boldsymbol{v}\cdot\boldsymbol{V}}{V}t}{\sqrt{\frac{V^2}{c^2}-1}},\nonumber \\
c\boldsymbol{t}' &=& \boldsymbol{v}t-\frac{\boldsymbol{v}\cdot\boldsymbol{V}}{V^2}\boldsymbol{V}t + \frac{\frac{\boldsymbol{v}\cdot\boldsymbol{V}}{Vc}-\frac{c}{V}}{\sqrt{\frac{V^2}{c^2}-1}}\boldsymbol{V}t
\end{eqnarray}
and eliminating $t$ from the equations we get:
\begin{align}
\label{eqmotionsup}
r' \frac{\sqrt{\frac{V^2}{c^2}-1}\left(\boldsymbol{v}-\frac{\boldsymbol{v}\cdot\boldsymbol{V}}{V^2}\boldsymbol{V}\right) 
- c\left(1-\frac{\boldsymbol{v}\cdot\boldsymbol{V}}{c^2} \right)\frac{\boldsymbol{V}}{V}}{V-\frac{\boldsymbol{v}\cdot\boldsymbol{V}}{V}} &= c\boldsymbol{t}'.
\end{align}
Now we can compare our result \eqref{eqmotionsup} with the equation \eqref{supervelocityprev} to find  the superluminal velocity addition formula:
\begin{align}
\label{superveltrans}
\frac{c^2}{v'^2} \boldsymbol{v'} = \frac{\sqrt{1-\frac{c^2}{V^2}}\left(\boldsymbol{v}-\frac{\boldsymbol{v}\cdot\boldsymbol{V}}{V^2}\boldsymbol{V}\right) 
- \left(\frac{c^2}{V^2} \boldsymbol{V}-\frac{\boldsymbol{v}\cdot\boldsymbol{V}}{V^2} \boldsymbol{V} \right)}{1-\frac{\boldsymbol{v}\cdot\boldsymbol{V}}{V^2}},
\end{align}
where $\boldsymbol{v'}$ is the velocity of the considered object in the superluminal frame of reference. Notice that the result \eqref{superveltrans} can be obtained directly from \eqref{veltrans} by substitution $\boldsymbol{V}\to \frac{c^2}{V}\frac{\boldsymbol{V}}{V}$ and $\boldsymbol{v'}\to \frac{c^2}{v'}\frac{\boldsymbol{v'}}{v'}$. Finally, we take a square of \eqref{superveltrans} to compute $v'^2$, plug it back and after some simplifications we find the velocity composition formula for a superluminal velocity $\boldsymbol{V}$, which is a superluminal version of \eqref{veltrans}:
\begin{widetext}
\begin{align}
\label{superveltrans2}
\boldsymbol{v'} &= 
\frac{\sqrt{1-\frac{c^2}{V^2} }\left(\boldsymbol{v}-\frac{\boldsymbol{v}\cdot\boldsymbol{V}}{V^2}\boldsymbol{V}\right) 
-\left(\frac{c^2}{V^2}\boldsymbol{V}-\frac{\boldsymbol{v}\cdot\boldsymbol{V}}{V^2} \boldsymbol{V} \right)}{1-\frac{\boldsymbol{v}\cdot\boldsymbol{V}}{V^2}} \left(1 - \frac{\left(1-\frac{c^2}{V^2}\right) \left(1-\frac{v^2}{c^2}\right)}{\left(1-\frac{\boldsymbol{v}\cdot\boldsymbol{V}}{V^2} \right)^2}\right)^{-1}.
\end{align}
\end{widetext}
By squaring the above result we also find the relation between speeds:
\begin{align}
\label{supermagnitudes}
\left(1-\frac{c^2}{v'^2}\right) = \frac{\left(1-\frac{c^2}{V^2}\right) \left(1-\frac{v^2}{c^2}\right)}{\left(1-\frac{\boldsymbol{v}\cdot\boldsymbol{V}}{V^2}\right)^2}.
\end{align}
The result \eqref{supermagnitudes} shows that if $v=c$ then $v'=c$. Therefore our superboost \eqref{super3d} preserves the speed of light. Moreover, if $v>c$ then $v'<c$ and vice versa. Therefore subluminal objects move with superluminal speeds according to any superluminal observer and superluminal objects according to any superluminal observer move with subluminal speeds. We must remember, however, that since both subluminal and superluminal families of observers are physically distinguishable, we can apply these adjectives to observers in an absolute sense. The constancy of the speed of light across all inertial frames is a property characteristic to the definition of velocity derived in \eqref{supervelocity}. Notice that alternative, incorrect definitions, such as one in \cite{Ziino1979} may lead to different conclusions.

The inverse transformation to \eqref{superveltrans} can be obtained either by using \eqref{superinverse3d} to evaluate $\boldsymbol{v}\equiv\frac{\text{d}\boldsymbol{r}}{\text{d}t}$ or by using \eqref{superveltrans} with the substitution: $\boldsymbol{V}\to-\boldsymbol{V}$ and $\frac{c^2}{v'^2}\boldsymbol{v'} \leftrightarrow \boldsymbol{v}$. The result is the same:
\begin{align}
\boldsymbol{v} &= 
\frac{\sqrt{1-\frac{c^2}{V^2} }\left(\frac{c^2\boldsymbol{v'}}{v'^2}-\frac{c^2}{v'^2}\frac{\boldsymbol{v'}\cdot\boldsymbol{V}}{V^2}\boldsymbol{V}\right) 
+\left(\frac{c^2\boldsymbol{V}}{V^2}+ \frac{c^2}{v'^2}\frac{\boldsymbol{v'}\cdot\boldsymbol{V}}{V^2} \boldsymbol{V} \right)}{1+\frac{c^2}{v'^2}\frac{\boldsymbol{v'}\cdot\boldsymbol{V}}{V^2}}.
\end{align}
Now, that the kinematics is well defined in superluminal frames of reference, and the corresponding transformation laws are established, let us discuss the expressions for energy and momentum of superluminal particles.

\section{Energy and momentum}

Let us begin by looking for the characterization of energy and momentum of a point-like particle moving in a subluminal frame with a superluminal velocity $\boldsymbol{v}$. As in conventional special relativity, we require that energy and momentum of a superluminal particle form a four-vector structure. This will guarantee that if energy and momentum are conserved for a certain process in one inertial frame, they will also be conserved for this process in all other reference frames. Such a process can include interactions between both sub- and superluminal objects, as well as light.

Similarly to conventional special relativity we will assume that the four-momentum $p^\mu$ of a superluminal particle is proportional to four-velocity $u^\mu$, but this time we will take the latter to be a space-like four-vector with $u^\mu u_\mu = -c^2$.

Consider a worldline of a superluminal object characterized by the four-velocity $u^\mu \equiv (u^0, \boldsymbol{u})$. We will look for this four-vector in a form analogous to the subluminal expression for which $\boldsymbol{u} = u^0\frac{\boldsymbol{v}}{c}$, so that $u^\mu = u^0 (1, \frac{\boldsymbol{v}}{c})$. Notice that the direction of velocity $\boldsymbol{v} = \frac{\text{d}\boldsymbol{r}}{\text{d}t}$ and the spacial component $\boldsymbol{u}$ are either parallel or antiparallel, depending on the sign of $u^0$. Since the four-vector $u^\mu$ is spacelike, even the subluminal Lorentz transformation \eqref{lorentz} can change the sign of $u^0$ and reverse the mutual orientation between $\boldsymbol{v}$ and $\boldsymbol{u}$. We require energy $E$ of the particle to be proportional to $u^0$ and momentum $\boldsymbol{p}$ proportional to $\boldsymbol{u}$, therefore we should be aware that the sign of energy will be frame dependent. Therefore a superluminal particle can either have positive energy and momentum parallel to velocity, or negative energy and momentum antiparallel to velocity in another frame of reference. This ambiguity results from the fact that specifying the mass and velocity of a superluminal object is insufficient to determine its energy and momentum. We also have to specify, whether the object in question is a particle, or an antiparticle moving backwards in time. Furthermore we should expect that in quantum field theory a tachyonic particle can be transformed into its antiparticle via a mere subluminal Lorentz transformation \eqref{lorentz}. This is essentially the same argument, as the one used by Feynman in his famous lecture on the relativistic reason for antiparticles \cite{Feynman1987}, in which he was focusing on superluminal virtual particles.

Let us compute the length of the superluminal four-velocity:
\begin{align}
\label{superunity}
c = \sqrt{
-u^\mu u_\mu
} &=
\sqrt{\boldsymbol{u}\cdot\boldsymbol{u}-\left(u^0\right)^2}
=|u^0|
\sqrt{\frac{v^2}{c^2}-1},
\end{align}
which gives:
\begin{align}
|u^0| = 
\frac{
c
}{
\sqrt{\frac{v^2}{c^2}-1}
},
\end{align}
that can be used together with $u^0 = \text{sgn}(u^0) |u^0|$ to define the the four-momentum in the following way:
\begin{align}
\label{fourenergy}
p^\mu \equiv m u^\mu = \text{sgn}(u^0)\left(\frac{mc\, }{\sqrt{\frac{v^2}{c^2}-1}}, \frac{m\boldsymbol{v}\, }{\sqrt{\frac{v^2}{c^2}-1}}\right),
\end{align}
where $m$ is the superluminal mass of the particle in analogy to subluminal particles. An identical procedure in another subluminal reference frame gives:
\begin{align}
\label{fourenergyprime}
p'^\mu = \text{sgn}(u'^0)\left(\frac{mc\, }{\sqrt{\frac{v'^2}{c^2}-1}}, \frac{m\boldsymbol{v'}\, }{\sqrt{\frac{v'^2}{c^2}-1}}\right)
\end{align}
and four-vectors \eqref{fourenergy} and \eqref{fourenergyprime} are related via a Lorentz boost \eqref{lorentz}. This relation for the temporal component yields:
\begin{align}
u'^0  &= \frac{u^0-\frac{\boldsymbol{u}\cdot\boldsymbol{V}}{c}}{\sqrt{1-\frac{V^2}{c^2}}},
\end{align}
which leads to the transformation properties of $\text{sgn}(u^0)$:
\begin{align}
\text{sgn}(u'^0) &=
\text{sgn}\left(
\frac{u^0-\frac{\boldsymbol{u}\cdot\boldsymbol{V}}{c}}{\sqrt{1-\frac{V^2}{c^2}}} \right) = \text{sgn}\left(u^0-\frac{\boldsymbol{u}\cdot\boldsymbol{V}}{c}
\right) \nonumber \\
&=
\text{sgn}(u^0)\,\text{sgn}\left(1-\frac{\boldsymbol{v}\cdot\boldsymbol{V}}{c^2}\right).
\end{align}
Finally, the relativistically covariant expressions for energy and momentum of a superluminal particle in subluminal reference frames are given by:
\begin{align}
E &\equiv \frac{\sigma mc^2}{\sqrt{\frac{v^2}{c^2}-1}},\nonumber \\
\boldsymbol{p} &\equiv \frac{\sigma m\boldsymbol{v}}{\sqrt{\frac{v^2}{c^2}-1}},
\end{align}
where $\sigma = \pm 1$ carrying the information whether the object is a particle or an antiparticle transforms according to the equation:
\begin{align}
\sigma' = \sigma\,
\text{sgn}
\left(
1-\frac{\boldsymbol{v}\cdot\boldsymbol{V}}{c^2}
\right),
\end{align}
which was first justified heuristically in \cite{Terletskii1968}. The important observation is that superluminal objects cannot be slowed down under the speed of light, as this would require infinite supplies of energy. Also notice that an object moving with an infinite velocity carries momentum, but no energy, as  $\lim_{v\to\infty}|\boldsymbol{p}| = mc$ and $\lim_{v\to\infty}E = 0$. 
This leads to an interesting novelty. In conventional relativity a particle cannot emit another particle without changing its own mass \cite{Dragan2021}. For example a free electron cannot emit a photon without violating conservation of energy. The situation is different if we include superluminal particles in our considerations. For example, a subluminal particle could emit an infinitely fast moving 
particle and simply reverse its own velocity due to recoil. In such a process, the energy is conserved, because the source does not change its energy and the emitted particle carries no energy. Momentum conservation can also be satisfied by an appropriate choice of the initial velocity or mass of the object. In principle, such an emission could be repeated a number of times and the object could emit a number of infinitely fast moving particles \cite{Sudarshan1962}.
Notice that such an emission is only possible after reaching certain velocity threshold. If velocity of the "source" is below the threshold, only the absorption of an infinitely moving particle is permitted unless one chooses to interpret such a process as an emission of an anti-particle backwards in time \cite{Sudarshan1962}.

The possibility of superluminal particles moving backwards in time was discussed by Bilaniuk and Sudarshan \cite{Sudarshan1969}. Since an ordinary Lorentz boost can transform a positive energy forward-in-time moving particle into its antiparticle (a negative energy, backwards-in-time moving particle), it is always possible to transform a problematic scenario with a backwards-in-time moving object to a frame, in which it moves forwards in time. When a negative-energy particle is emitted from a source backwards in time, it is always possible to “reinterpret”  such a scenario as the one, in which a positive-energy particle is moving forwards in time and being absorbed by the source. Probability amplitudes computed in quantum field theory remain invariant under such a “reinterpretation”, which was pointed out in \cite{Sudarshan1969}. Also, Dragan and Ekert in \cite{Dragan2020}  argued that no explicit causal paradoxes result from such a reinterpretation as long as the classical notion of determinism is abandoned.

Energy and momentum can also be transformed to superluminal frames of reference. Similarly to four-position, the four-momentum in these frames contains the energy three-vector and a single momentum. Applying the superboost \eqref{super3d} to the four-vector \eqref{fourenergy} instead of four-position $(ct, \boldsymbol{r})$ we obtain:
\begin{eqnarray}
p' &=& \frac{\frac{VE}{c^2}-\frac{\boldsymbol{p}\cdot\boldsymbol{V}}{V}}{\sqrt{\frac{V^2}{c^2}-1}},\nonumber \\
\frac{\boldsymbol{E}'}{c} &=& \boldsymbol{p}-\frac{\boldsymbol{p}\cdot\boldsymbol{V}}{V^2}\boldsymbol{V} + \frac{\frac{\boldsymbol{p}\cdot\boldsymbol{V}}{Vc}-\frac{E}{Vc}}{\sqrt{\frac{V^2}{c^2}-1}}\boldsymbol{V}.
\end{eqnarray}
Transformation to the rest frame of the superluminal particle, as previously, can also be decomposed into two steps: first transforming the four-momentum to a frame, in which the object moves with an infinite velocity in some direction $\boldsymbol{s}$ and then applying the analogue of the superflip \eqref{inf3d} that interchanges energy with momentum. This results in the four-momentum taking the form:
\begin{align}
p'^\mu = (mc\boldsymbol{s}, 0),	
\end{align}
which is not unique and corresponds to a bundle of infinitely many worldlines for each choice of $\boldsymbol{s}$. This shows that there are infinitely many observers characterized by non-parallel worldlines, each given by a different $\boldsymbol{s}$, that are at relative rest. To confirm this it suffices, to inspect the definition of velocity in superluminal frames \eqref{supervelocity} that shows that if $\text{d}r'=0$ then the velocity vanishes. This interesting conundrum will be clarified once we turn our attention to the dynamical laws including superluminal extension of relativity.

\section{Dynamics}

In classical mechanics, a dynamical evolution of a system is obtained by minimizing the action: 
\begin{equation}
\label{subaction}
S \equiv \int L\, \text{d}t.	
\end{equation}
Frame-independence of the result is guaranteed by demanding $S$ to be the same in all (subluminal) frames and for a single point-like particle the result is a one-dimensional trajectory $\boldsymbol{r}(t)$. Now we wish to generalize this principle of least action to include superluminal frames. As we now have to deal with three time variables, $\boldsymbol{t}'$, we hypothesize that the corresponding variational principle will involve an action of the form:
\begin{equation}
\label{superaction}
S' \equiv \int L'\, \text{d}^3t',	
\end{equation}
where $L'$ is an appropriately defined superluminal counterpart of the Lagrangian. Clearly, the resulting "trajectory" of a single particle $r'(\boldsymbol{t'})$ defines a three-dimensional surface, which is apparently at odds with subluminal intuitions. It is therefore interesting to explore this notion in more detail.

We are seeking to write down $L'$ such that $S'$ is relativistically invariant. With one time variable,
it would be accomplished with $L$ proportional to the length
of the one-dimensional trajectory $\boldsymbol{r}(t)$, i.e. we would take 
\begin{equation}
    L\propto \sqrt{1-\frac{1}{c^2}\left(\frac{\mathrm{d}\boldsymbol{r}}{\mathrm{d}t}\right)^2}.
\end{equation}
Similarly, with three time variables, the natural choice would be to consider the three-dimensional volume of the manifold parametrized as $r(\boldsymbol{t})$, i.e.\ to consider
\begin{equation}
L'\propto \sqrt{\frac{1}{c^2}\left( \boldsymbol{\nabla_{\!t'}} r'\right)^2-1},
\label{kteq:action3}
\end{equation}
where $\boldsymbol{\nabla_{\!t'}}$ denotes a gradient with respect to the time variables. The choice of signs in the square root reflects the fact that we are interested in motions which are superluminal in the chosen reference frame, i.e. we characterize the example of subluminal particles in a superluminal frame. With such a choice of $L'$, the action (\ref{superaction}) is evaluated over $\mathbb{R}^3$ of the time variables. This is strikingly different from a one-dimensional case, in which instants of time can be ordered, so we can evaluate the action (\ref{subaction}) over a finite interval.

Demanding that the action (\ref{superaction}) is stationary with respect to the variations of the manifold $r'(\boldsymbol{t'})\to r'(\boldsymbol{t}')+\delta r'(\boldsymbol{t'})$ leads to:
\begin{widetext}
\begin{align}
0 = \delta S' &\propto \int \frac{\boldsymbol{\nabla_{\!t'}} r'\cdot \boldsymbol{\nabla_{\!t'}} \delta r'}{\sqrt{\frac{1}{c^2}\left( \boldsymbol{\nabla_{\!t'}} r'\right)^2-1}}\,\mathrm{d}t'^3
=\int \frac{\left( \left( \frac{1}{c^2}(\boldsymbol{\nabla_{\!t'}}r')^2-1\right)\delta_{ij}-\frac{1}{c^2}\frac{\partial r'}{\partial t'_i}\frac{\partial r'}{\partial t'_j}\right) \frac{\partial^2 r'}{\partial t'_i \partial t'_j}}{\left(\frac{1}{c^2}\left( \boldsymbol{\nabla_{\!t'}} r'\right)^2-1\right)^{3/2}}\,\delta r'\,\mathrm{d}t'^3,
\end{align}
\end{widetext}
where in the last equality we assumed that the boundary term vanishes at infinity and we can evaluate the integral by parts; summation over repeated indices is understood. Because $\delta r'$ is an arbitrary function of $\boldsymbol{t'}$, we conclude that the stationary action corresponds to the equation of motion
\begin{equation}
\left( \left( \frac{1}{c^2}(\boldsymbol{\nabla_{\!t'}}r')^2-1\right)\delta_{ij}-\frac{1}{c^2}\frac{\partial r'}{\partial t'_i}\frac{\partial r'}{\partial t'_j}\right) \frac{\partial^2 r'}{\partial t'_i \partial t'_j} = 0.
\label{kteq:laplace}
\end{equation}

Because \eqref{kteq:laplace} contains second-order time derivatives,
it is solved by functions of the form:
\begin{equation}
r'(\boldsymbol{t'})=r'_0+\boldsymbol{v'}\cdot \boldsymbol{t}',    
\label{kteq:lin}
\end{equation}
where $r'_0$ and $\boldsymbol{v'}$ are constant. Those solutions, that are compatible with the equation of constant motion \eqref{velproduct}, can be viewed as "sheets" of particles traveling with velocity $\boldsymbol{v'}$ perpendicular to the sheets. This behavior corresponds to the motion of a collection of point-like particles described in a subluminal frame. Unfortunately, this is where the similarities of the two points of view end. In a spherically symmetric case, eq.~\eqref{kteq:laplace} becomes:
\begin{equation}
    \frac{\mathrm{d}^2r'}{\mathrm{d}{t'}^2} + \frac{2}{t'} \frac{\mathrm{d}r'}{\mathrm{d}{t'}}- \frac{2}{c^2 t'}\left( \frac{\mathrm{d}r'}{\mathrm{d}{t'}} \right)^3
 = 0 \, ,
 \label{kteq:laplace_sph}
\end{equation}
where $t'\equiv|\boldsymbol{t'}|$. Solutions of eq.~\eqref{kteq:laplace_sph} include a constant,
a spherical shell moving with the speed of light and in other cases they are expressed in terms of elliptic functions. In particular, the latter solution does not correspond to any known representation of a free particle or a collection thereof.

The problems that we encountered after swapping the temporal and spatial coordinates by means of a superluminal transformation follow from the dimensionality of space. Were there only one spatial dimension, the partial differential equations involving $\boldsymbol{\nabla_{\!t'}}$ would become ordinary differential equations and linear trajectories of point-like particles would remain linear trajectories. 

Furthermore, the choice of \eqref{kteq:action3} is not unique. However, any action that respects the shift and rotational symmetry of $\boldsymbol{t}'$ will depend only on $(\boldsymbol{\nabla_{\!t'}} r)^2$, which will lead to the second order differential equation like \eqref{kteq:laplace} for a free particle. In this sense, the derivation presented above is quite general, as \eqref{kteq:lin} would still be the solution of 
the resulting equation of motion.

What is even worse, the formalism described above gives us no chances to arrive at a relation between the actions \eqref{subaction} and \eqref{superaction} as their solutions are clearly non-equivalent. We are therefore forced to conclude that our approach of treating $r'(\boldsymbol{t}')$ as the parametrization of the configuration space of the system fails badly on many fronts.

Let us take a step back and try to formulate our requirements for a satisfactory description of a physical system which is relativistically invariant and, in particular, does not break down for a superluminal transformation. To this end, let us symbolically denote the state of the system fully characterizing it in a configuration space in a subluminal frame by $\psi$ and similarly in the superluminal frame by $\psi'$. We will also assume that the transition to the superluminal frame transforms $\psi \to \psi'$ so that the action $S$ given by \eqref{subaction} can be written down in both subluminal and superluminal frames as:
\begin{align}
\label{subaction2}
S \equiv \int L[\psi]\text{d}t = \int L[\psi']\frac{\text{d}r'}{c},
\end{align}
where in the last equality we applied the superflip \eqref{inf3d}. We also have similar relations for the superluminal action $S'$ given by \eqref{superaction}:
\begin{align}
\label{superaction2}
S' \equiv \int L'[\psi']\text{d}^3t' = \int L'[\psi]\frac{\text{d}^3r}{c^3}.
\end{align}
In order to guarantee that the principle of least action leads to equivalent solutions in all frames we will enforce that the actions \eqref{subaction2} and \eqref{superaction2} are equal: $S=S'$. This leads to the following equalities:
\begin{align}
\label{subeqsuper1}
&\int L[\psi]\text{d}t = \int L'[\psi]\frac{\text{d}^3r}{c^3},
\end{align}
and
\begin{align}
\label{subeqsuper2}
&\int L[\psi']\frac{\text{d}r'}{c} =\int L'[\psi']\text{d}^3t'.
\end{align}
It follows that the characterization of the state in terms of a "trajectory" is no longer possible, because the Lagrangian $L'$ must now be integrated over $\text{d}^3r$ and the Lagrangian $L$ must be integrated over $\text{d}r'$, therefore $\psi \neq \boldsymbol{r}(t)$ and $\psi' \neq r'(\boldsymbol{t'})$. This circumstance brings us to the conclusion that it is necessary to introduce a new type of configuration space and a new notion of a state of the system, $\psi(t, \boldsymbol{r})$, that depends on all spacetime coordinates, over which the action will be minimized. For such states, the function encoding the state of the
system has to be integrated over both time and space, hence it should
be interpreted as a Lagrangian density of the system.
Of course, such a physical picture brings us outside of the domain of classical mechanics and forces us to consider a familiar framework involving fields as the only possible descriptions of physical systems.

Let us now observe that if the Lagrangian $L$ on the left-hand side of \eqref{subeqsuper1} only involves integration over $\text{d}t$, while $L'$ on the right-hand side over $\text{d}^3r$, we would face the equality of a pair of functions of different variables: $\boldsymbol{r}$ on the left-hand side and $t$ on the right-hand side. In order to avoid this without introducing preference for any of the spacetime point, we are forced to impose that the Lagrangians must be expressed in terms of Lagrangian densities: $L[\psi]\equiv\int{\cal L} [\psi]\,\text{d}^3 r$ and $L'[\psi']\equiv\int{\cal L}[\psi'] \,\text{d} r'$, so that the additional variables are also integrated over. This also solves an analogous difficulty with equation \eqref{subeqsuper2}. All in all, we find that extending the special relativistic dynamics to superluminal observers terminates any chance to consider mechanical systems and forces one to consider a field-theoretic configuration space. In order to specify whether the resulting "field theory" should be classical or quantum one has to refer to arguments put forth in \cite{Dragan2020} that ruled out a conventional deterministic theory. Whether the resulting theory can only be quantum field theory or other alternatives are also possible \cite{bohm2006undivided,Nikolic,Foo2022} is yet to be determined.
What remains out of question is that the superluminal extension of special relativity in $1+3$ spacetime is possible in a fully consistent way. However, for this task to be covariant it is necessary to abandon the classical picture of mechanical systems characterized by unique, one-dimensional trajectories and upgrade physics to the field-theoretic framework.

\section{Example: Maxwell's theory}

In Maxwell's theory the state of the system is characterized by electromagnetic potentials and their derivatives: $\psi\to\{A^\mu, \partial_\nu A^\mu \}$ and we have the following Lagrangian density in Lorenz gauge $\partial_\mu A^\mu = 0$:
\begin{align}
\label{LangDens}
{\cal L}(A^\mu) \equiv -\frac{1}{4\mu_0}F^{\mu\nu}F_{\mu\nu}- A^\mu j_\mu,
\end{align}
where $F_{\mu\nu} \equiv \partial_\mu A_\nu - \partial_\nu A_\mu$ and $\partial_\mu \equiv (\frac{1}{c}\partial_t,\boldsymbol{\nabla})$. The resulting Euler-Lagrange equations are:
\begin{align}
\label{ELeqs}
\partial^\mu\partial_\mu A^\nu = \mu_0 j^\nu.
\end{align}
Let us now formulate the theory in superluminal reference frames. This problem has already been addressed by Dawe and Hines \cite{DaweIII,DaweIV}, however they adopted the interpretation in which superluminal observers are characterized by a single temporal dimension and non-Euclidean space \cite{DaweI}. Our approach, assuming the three-dimensional time variable, provides a much cleaner mathematical framework and the task of deriving field equations becomes practically trivial. Since an arbitrary superluminal transformation \eqref{super3d} can be decomposed into a subluminal boost (which leaves Maxwell's equations intact) and the infinite velocity superflip \eqref{inf3d}, it is sufficient to only consider the latter.

The superflip \eqref{inf3d} imposes that $\partial'_\mu = (\frac{1}{c}\boldsymbol{\nabla_{\!t'}}, \partial_{r'})$, as well as the following set of conditions for the four-potentials $A^\mu$ and four-current $j^\mu$:
\begin{align}
A'^\mu &\equiv (\frac{1}{c}\boldsymbol{\varphi'}, A') = (\boldsymbol{A},\frac{1}{c}\varphi),\nonumber \\
j'^\mu &\equiv (\boldsymbol{\varrho'}c, j') = (\boldsymbol{j},\varrho c).
\end{align}
The transformation procedure characterized in the previous Section requires that the Lagrangian density \eqref{LangDens} and dynamical equations \eqref{ELeqs} in the superluminal frame have the analogous form:
\begin{align}
\label{LangDensSup}
{\cal L}(A'^\mu) = -\frac{1}{4\mu_0}F'^{\mu\nu}F'_{\mu\nu}+ A'^\mu j'_\mu,
\end{align}
where $F'_{\mu\nu} \equiv \partial'_\mu A'_\nu - \partial'_\nu A'_\mu$ and the sign change in the source term is due to the equation \eqref{intervaltrans}, so that $\eta_{\mu\nu}A^\mu j^\nu = - \eta'_{\mu\nu}A'^\mu j'^\nu $. The resulting dynamical equations are
\begin{align}
\label{ELeqsSup}
-\partial'^\mu\partial'_\mu A'^\nu = \mu_0 j'^\nu,
\end{align}
which can be also obtained directly by applying the superflip \eqref{inf3d} to \eqref{ELeqs}. Our construction of electromagnetic potentials in superluminal frames is manifestly covariant. We can also introduce electric and magnetic fields in subluminal frames in a standard way:
\begin{align}
\boldsymbol{E} &\equiv -\boldsymbol{\nabla}\varphi - \partial_t\boldsymbol{A}, \nonumber \\
\boldsymbol{B} &\equiv \boldsymbol{\nabla} \times \boldsymbol{A}
\end{align}
and define these fields in superluminal frames using the following analogies:
\begin{align}
\boldsymbol{E'} &\equiv -\boldsymbol{\nabla_{\!t'}}A' - \partial_{r'}\boldsymbol{\varphi'}, \nonumber \\
\boldsymbol{B'} &\equiv \boldsymbol{\nabla_{\!t'}} \times \boldsymbol{\varphi'}.
\end{align}
Then the Maxwell's equations in superluminal frames take the form:
\begin{align}
\label{superMax}
\boldsymbol{\nabla_{\!t'}}\cdot\boldsymbol{E'} &= -\frac{1}{\varepsilon_0 c} j'\nonumber \\    
\boldsymbol{\nabla_{\!t'}}\cdot\boldsymbol{B'} &= 0 \nonumber \\    
\boldsymbol{\nabla_{\!t'}}\times\boldsymbol{E'} &= -\partial_{r'}\boldsymbol{B'}\nonumber \\    
\boldsymbol{\nabla_{\!t'}}\times\boldsymbol{B'} &= -\mu_0 c\boldsymbol{\varrho'} + \frac{1}{c}\partial_{r'}\boldsymbol{E'}.
\end{align}
In the special case of static fields, the Amp\`ere law given by the last equation \eqref{superMax} reduces to the superluminal form of the "Gauss law":
\begin{align}
\partial_{r'}\boldsymbol{E'} = \frac{\boldsymbol{\varrho'}}{\varepsilon_0}.
\end{align}
Our four-dimensional formalism introduced in this work makes the derivation of the superluminal form of Maxwell's equations almost effortless. The lack of invariance of Maxwell's theory under superboosts is a signature of the fact that superluminal observers physically differ from their classical counterpart.

\section{Conclusions}

We have explicitly shown how to extend special relativity to allow superluminal inertial observers in $1+3$ dimensional spacetime. Admittedly, these observers can be physically distinguished from the subluminal ones, but we show that they are indistinguishable among themselves, just like all the subluminal inertial observers are. One of the most challenging and counter-intuitive aspects of our construction is the fact that spacetime metric in superluminal frames transforms itself into a $3+1$ dimensional one. Therefore we developed the whole kinematics starting with a derivation of a sensible and useful definition of velocity in a spacetime involving three-dimensional time and one-dimensional space. We showed that the speed of light is still preserved by the superboosts, therefore the task of finding all possible transformations preserving the speed of light may not be considered complete with just conventional Lorentz boosts. A subtlety that is easy to overlook is that superluminal observers require their own, unorthodox definition of velocity. Furthermore, we characterized expressions for energy and momentum of superluminal particles and discussed some of their novel properties, absent in conventional subluminal special relativity.

Finally, we showed that dynamical theories based on a relativistically invariant principle of least action including superluminal observers cannot be based on a mechanical paradigm of a classical, point-like trajectory. Inclusion of superluminal observers leads to a principle of least action based on a field-theoretical framework with a notable example of the Maxwell's theory. Therefore one of the most interesting aspects of extending special relativity to superluminal frames of reference is the emergence of field theory. The need for a field-theoretic description of matter in terms of "matter waves" came out of the blue in the early era of quantum theory. Until recently it was generally believed that postulates underlying quantum theory are fundamental and cannot be derived from anything more basic. In this work we showed that the justification of quantum theory using extended relativity, presented in \cite{Dragan2020}, can be naturally generalized to $1+3$ spacetime and such an extension leads to the field-theoretic formulation of the quantum theory. This justifies, or at least provides a plausibility argument, why this extension is not just an eccentric thought exercise, but reflects something fundamental about symmetries of laws of physics.

A relevant question that remains to be answered is whether tachyons -- understood as localized lumps of energy traveling with superluminal speeds -- can physically exist \cite{Ehrlich2022}. As was shown here, and earlier in \cite{Dragan2020}, special relativity does not exclude such a possibility. In quantum field theory, mechanisms of spontaneous symmetry breaking, such as the Higgs mechanism, involve fields whose mass squared is negative in the unbroken phase \cite{Srednicki}, so excitations of those fields can be regarded as tachyons \cite{Feinberg, Sudarshan1968}. In the phase with broken symmetry, those field are expanded around a degenerate local minimum, so their excitations have positive mass squared and are known as Higgs particles behaving as regular, subluminal particles. However the basic principle of the mechanism always involves tachyons as a starting point. Given rather interesting kinematics and dynamics of superluminal particles discussed in this work, it would be interesting to explore this initial phase before symmetry breaking in more detail and the role of tachyons therein. We believe our current work should stimulate new research in this direction.

\begin{acknowledgments}
We would like to thank Iwo Białynicki-Birula and Tomasz Miller for useful comments.

K.~D.\ is financially supported by the (Polish) National Science Center Grant 2021/41/N/ST2/01901. K.~T.\ is partially supported by grant 2018/30/Q/ST9/00795 from the (Polish) National Science Centre.
\end{acknowledgments}

\bibliography{library}

\end{document}